# Photonic transport control by spin-optical disordered metasurface

Dekel Veksler, Elhanan Maguid, Dror Ozeri, Nir Shitrit, Vladimir Kleiner, and Erez Hasman[*]

Micro and Nanooptics Laboratory, Faculty of Mechanical Engineering, and Russell Berrie Nanotechnology Institute, Technion – Israel Institute of Technology, Haifa 32000, Israel

[*]e-mail: mehasman@technion.ac.il

**Photonic metasurfaces are ultrathin electromagnetic wave-molding metamaterials[1-4] providing the missing link for the integration of nanophotonic chips with nanoelectronic circuits. An extra twist in this field originates from spin-optical metasurfaces providing the photon spin (polarization helicity) as an additional degree of freedom in light-matter interactions at the nanoscale[4-6]. Here we report on a generic concept to control the photonic transport by disordered (random) metasurfaces with a custom-tailored geometric phase. This approach combines the peculiarity of random patterns to support extraordinary information capacity within the intrinsic limit of speckle noise, and the optical spin control in the geometric phase mechanism, simply implemented in two-dimensional structured matter. By manipulating the local orientations of anisotropic optical nanoantennas, we observe spin-dependent near-field and free-space open channels, generating state-of-the-art multiplexing and interconnects. Spin-optical disordered metasurfaces provide a route for multitask wavefront shaping via a single ultrathin nanoscale photonic device.**

The ability to control the flow of light beyond that offered by conventional optics has significantly improved owing to the rapidly expanding field of photonic metasurfaces[1-4]. Metasurfaces are two-dimensional metamaterials composed of engineered subwavelength-scale meta-atoms enabling the manipulation of an incident electromagnetic wave by an abrupt change of the phase over a subwavelength distance. By molding the polarization[7], linear and angular momenta[5,6,8], custom-tailored metasurfaces have been utilized for ultrathin planar optical devices[3,9,10]. Metasurfaces are realized in either dielectric-dielectric[2,9] or metal-dielectric interfaces[4-8,10,11], where in the latter, propagating surface-confined waves of surface plasmon polaritons (SPPs – resonant collective oscillations of quasi-free electrons at the metal surface[12]) mediate the in- and out-coupling of light.

The Pancharatnam-Berry phase[13] is a promising approach for achieving an abrupt phase change leveraging the design of metasurfaces, as originally presented in ultrathin metallic[1] and dielectric phase optical elements[2]. The peculiarity of this phase lies in its geometric nature; unlike diffractive and refractive elements, it does not arise from optical path differences but from a space-variant manipulation of the light polarization state[1,2,5,6,9]. When an incident circularly polarized light is scattered from a metasurface consisting of subwavelength anisotropic antennas whose local orientation angle is $\theta$, a geometric phase shift of $2\sigma\theta$ is induced, where $\sigma_{\pm} = \pm 1$ is the photon spin corresponding to right and left circular polarizations, respectively[2,5,6]. Apparently, the optical spin provides an additional degree of freedom in nano-optics for spin degeneracy removal phenomena in metasurfaces[4-6], such as polarization-controlled directional excitation of SPPs[14-17]. However, the presented unidirectional launching[14-16] via periodic metasurfaces suffers from the limitation of a single channel; moreover, the observed multidirectional excitation[17] reveals the constraints of a directional dependence between channels arising from the lattice symmetry, and limited number of channels due to rotational symmetry restrictions of a periodic metasurface crystal. These disadvantages conflict with the growing demand of multiple channels with independent directions and wavelengths for multifunctional metasurface devices. We offer to overcome the limitations of polarization-controlled directional excitation and thus expanding its scope via disordered metasurfaces with a custom-tailored geometric phase.

Previous research in the field of metasurfaces focused on ordered structures. Disordered systems were rarely addressed, despite increasing scientific efforts in the field of disordered photonics[18,19]. The exploration of new phenomena in random metasurfaces by coherent bulk effects, such as backscattering[20] and enhanced transmission through coupling to eigenchannels[21,22], has inspired even greater interest. Recently, phenomena related to random metasurfaces have been under investigation, among them are broadband and wide angle absorbers for solar cells[23], localized electromagnetic fields[24,25] and second-harmonic generation[26]. Here, we report on a novel generic concept to control photonic transport exploiting the peculiarities of disordered metasurfaces. We utilize the geometric phase, induced by the nanoantenna orientation degree of freedom, to open discrete spin-dependent channels in metasurfaces with randomly distributed antennas in both near (see Fig. 1a) and far fields (see Fig. 3a). Moreover, the revealed extraordinary channel capacity in disordered metastructures ushers in spin-optical multiplexing and interconnect metasurface devices.

The scattering from an isotropic nanohole, excited by circularly polarized light, results in a propagating SPP wave acquiring an orbital angular momentum (AM) that is equal to the incident spin AM[27]. The arising spin-based plasmonic electric field in polar coordinates $(r,\varphi)$ is $E(\sigma) \propto e^{i(kr-\sigma\varphi)}/\sqrt{r}$, where $\sigma$ is the incident optical spin, $k(\omega)$ is the SPP wave number, and $\omega$ is the frequency of light.

The AM conservation in this light-matter interaction originates from the circular symmetry of the scatterer. However, when an anisotropic nanoantenna is considered as a source, the AM is not conserved and an additional wave with an opposite orbital AM is generated (see Supplementary Section 1 for a detailed analysis), so

$$E(\sigma) \propto \left(e^{i(kr-\sigma\varphi)} + e^{i(kr+\sigma\varphi-2\sigma\theta)}\right)/\sqrt{r}. \tag{1}$$

The secondary surface wave is accompanied by a geometric phase of $-2\sigma\theta$ resulting from an AM conversion (Supplementary Section 1). By considering a metasurface consisting of an ensemble of uncoupled nanoantennas, the global SPP field is the coherent superposition of all the elemental fields. At an observation point far from the ensemble, the emerged field is $E(\mathbf{k},\sigma) \propto e^{-i\sigma\bar{\varphi}}\sum_n^N e^{-i\mathbf{k}\cdot\mathbf{r}_n} + e^{i\sigma\bar{\varphi}}\sum_n^N e^{-i(\mathbf{k}\cdot\mathbf{r}_n+2\sigma\theta_n)}$, where $\bar{\varphi}$ is the mean azimuthal angle of observation, $N$ is the total number of antennas per channel, and $\mathbf{r}_n$ and $\theta_n$ are the position vector and the orientation of the $n$th antenna, respectively.

The scattered field component $\sum_n^N e^{-i(\mathbf{k}\cdot\mathbf{r}_n+2\sigma\theta_n)}$ can be regarded as the structure factor of a metasurface, whereas the geometric phase is the spin-dependent atomic form factor of a nanoantenna[28]. Although a random distribution of the nanoantenna locations is considered, a proper selection of the antenna orientations results in a constructive interference when the phase-matching condition $\mathbf{k}\cdot\mathbf{r}_n + 2\sigma\theta_n = 2\pi m$ is fulfilled for an arbitrary integer $m$. We regard anisotropic antennas of nanorods whose local orientation is mod $\pi$ defined; hence, the above condition is reduced to

$$2\theta_n = \mathbf{k}_g \cdot \mathbf{r}_n, \tag{2}$$

whereas $\mathbf{k}_c = -\sigma\mathbf{k}_g$ is the desired channel wave vector. Accordingly, a spin-controlled channel is opened in a predetermined direction of $\mp\hat{\mathbf{k}}_c$ for $\sigma_\pm$ excitations, respectively. The sums in the scattered field $\sum_n^N e^{-i\mathbf{k}\cdot\mathbf{r}_n}$ and $\sum_n^N e^{-i(\mathbf{k}-\mathbf{k}_c)\cdot\mathbf{r}_n}$ can be evaluated by Monte Carlo integration theory[29]. This results in a zero (ballistic) diffraction order and the desired open channel, respectively, accompanied by a speckle noise $\varepsilon(\mathbf{k}) \propto O(\sqrt{N})$, i.e., a random distribution resulting from the coherent interference of wavefronts scattered from a fine-scale granular pattern[30]. Consequently, the total scattered field arises in a spin-dependent open channel expressed as $E(\mathbf{k},\sigma) \propto N\Delta((\mathbf{k}+\sigma\mathbf{k}_g)D) + \varepsilon(\mathbf{k})$, where $\Delta(\mathbf{p}) \equiv \mathrm{sinc}(p_x)\mathrm{sinc}(p_y)$, and $D$ is the metasurface width.

The introduced concept enables the design of multiple directional channels by a random mixing of antennas with different orientation functions in a single metasurface, where each channel is individually controlled by the wavelength and the polarization helicity of the incident light (see Fig. 1b). The number of open channels $N_c$ is restricted by the characteristic distance between antennas in each channel $d \approx \sqrt{A/N}$ and the minimal separation between neighboring antennas of $r_{min} \approx \sqrt{A/N_t}$, determined by fabrication limitations and the requirement for eliminating the coupling between nanoantennas[27]. Here, $A$ is the metasurface area and $N_t = NN_c$ is the total number of antennas. Accordingly, the geometric limit for the metasurface channel capacity is $N_c^{(g)} \approx (d/r_{min})^2$. For a random distribution of the antenna positions, there is no restriction on the distance $d$ for opening a single channel. On the other hand, for an ordered (periodic) antenna distribution, the scattered field consists of diffraction orders as manifested by the momentum-matching condition $\mathbf{k}_c = -\sigma\mathbf{k}_g + i\mathbf{G}_1 + j\mathbf{G}_2$, where $(\mathbf{G}_1,\mathbf{G}_2) = 2\pi/d(\hat{\mathbf{x}},\hat{\mathbf{y}})$ are the reciprocal lattice vectors. Here, a single channel ($i=j=0$) is obtained only for a subwavelength structure with $d < \lambda_c/2$, where $\lambda_c = 2\pi/k_c$. This implies that disordered metasurfaces provide an enormous advantage compared to ordered systems as they enable a sampling of the desired phase profile with $d >> \lambda_c/2$, which is essential for opening multiple channels under the geometric limitation of $N_c^{(g)}$ (Supplementary Section 3).

Plasmonic disordered metasurfaces were realized for the experimental observation of open channels in the near field. For each chosen channel, we control the propagation direction of the SPPs, launched by an array consisting of anisotropic void nanoantennas at random locations, by tuning the local antenna orientation (equation (2)). We implemented two spin-dependent channels operating at different incident wavelengths. Accordingly, the total number of antennas was randomly divided into two equal mixed groups. The first antenna group opens a spin-based SPP channel at the wavelength of 740 nm in 0° and 180° directions for $\sigma_\pm$ (Fig. 1e,f), respectively, whereas the second group opens a channel at 800 nm in 250° and 70° for $\sigma_\pm$, respectively (Fig. 1c,d). The fabricated metasurface was surrounded by an annular decoupling slit enabling free-space imaging of SPP jets launched from the antenna array. The metasurface was normally illuminated with a continuous wave Ti-sapphire tunable laser via a circular polarizer. The spin-controlled multichannel excitation was observed by measuring the intensity distributions along the slit (Fig. 1c-f). These spin-based open channels via a disordered metasurface were verified by calculated SPP intensity distributions (see Supplementary Section 2), obtained by the superposition of scattered fields from anisotropic antennas with designed orientations (equation (1)). Note that when the antenna orientations are randomly set, open channels are not observed (Fig. 1g).

Several critical issues arise when characterizing the near-field information capacity of a disordered multichannel metasurface: diffraction, noise, crosstalk and geometry. From a diffraction limit consideration, the upper limit for the number of channels is $N_c^{(d)} \approx 2\pi D / \lambda_{SPP}$, where $\lambda_{SPP}$ is the SPP wavelength. Alongside, the origin of the system noise is the speckle pattern, and the crosstalk between channels. The signal-to-noise ratio (SNR), which determines the number of bits per channel of $\log_2(1+\text{SNR})$, provides a limit for the channel capacity of $N_c^{(n)}$, set by $\text{SNR} \approx 1$. In a metasurface designed to open two spin-dependent channels, the $\sigma_+$ channels launch SPP jets in 0° and 180° directions for 740 and 800 nm, respectively (Fig. 2b,d), whereas $\sigma_-$ channels are oppositely directed (Fig. 2a,c). The measured azimuthal cross sections show that the crosstalk between the two open channels is rather weak (Fig. 2a-d). Regarding the dependence of the signal intensity in the open channel number, we investigated an additional metasurface with a single spin-controlled channel opened at 740 nm in 0° and 180° directions for $\sigma_\pm$ excitations, respectively. A ratio of four between the measured jet intensities from the single- and double-channel metasurfaces was observed (Fig. 2e). This coincides with a calculation based on the interference model (Supplementary Section 4) showing that the intensity of each channel scales as $(N_t/N_c)^2$ (Fig. 2f), as originally introduced in multiplex holograms[31]. Moreover, the calculation reveals that the noise intensity scales as $N_t$, so the SNR is proportional to $N_t/N_c^2$ (Fig. 2g), as experimentally observed in the single- and double-channel metasurfaces. For a chosen metasurface with a relatively small area of $10 \times 10 \, \mu m^2$, we obtain a channel capacity upper limit of $N_c^{(d)} \approx 90$, whereas the actual limitation of $N_c^{(n)} \approx 30$ (Fig. 2g), offering a multichannel design of tens of open channels.

Disordered metasurfaces can also open channels in the far field providing the route for free-space interconnects. Optical interconnects offer low crosstalk, high bandwidth and parallel operation, making them attractive for analog and digital optical computing as well as for electronic chips[32]. We demonstrated a general fan-out interconnect based on an ultrathin spin-optical disordered metasurface (Fig. 3a). By orientating the nanoantennas of each channel according to equation (2), a spin-controlled free-space channel with a transverse momentum shift is opened. The corresponding scattered component undergoes deflection at an angle of $\arcsin(\sigma k_g / k_0)$, and spin flip to an opposite spin state with regard to the incident beam with the wave number $k_0$[2,5,6]. The peculiarity of the disordered approach enabling to open channels with $d > \lambda/2$ was presented via a metasurface wherein $d \approx 2\lambda$ (Fig. 3d,e). We also observed 3x2 spin-dependent channels in desired directions (Fig. 3b,c), thereby introducing the ability to utilize light control by disordered metasurfaces for interconnects. More generally, spin-optical disordered metasurfaces with mixed groups of nanoantennas pave the way for controlling the light transport with a different beam-shaping task for each group.

The channel capacity of free-space optical interconnects based on metasurfaces with mixed antenna groups can be analyzed by the Gabor theory of information[33]. Multiple channels can be opened by two types of metasurfaces wherein the area is divided into $N_c$ separated regions (Fig. 3f), or the nanoantennas of each channel are randomly distributed over the entire area (Fig. 3g). For a given solid angle of $\Omega$, the diffraction limit of each channel states that for the first type $N_c \approx \sqrt{A\Omega/\lambda^2}$, whereas for the mixed channel type, according to Gabor limit, $N_c \approx A\Omega/\lambda^2$ [33,34]. Consequently, the information capacity of the presented disordered metasurface is significantly higher than the separated channel regions (see Fig. 3f,g). The reported concept provides the route for controlling the propagation direction of electromagnetic waves via state-of-the-art spin-optical nanoscale devices which can integrate with nanoelectronic circuits, ushering in a new era of light manipulation.

**Acknowledgements**

This research was supported by the Israel Science Foundation and the Israel Nanotechnology Focal Technology Area on Nanophotonics for Detection.

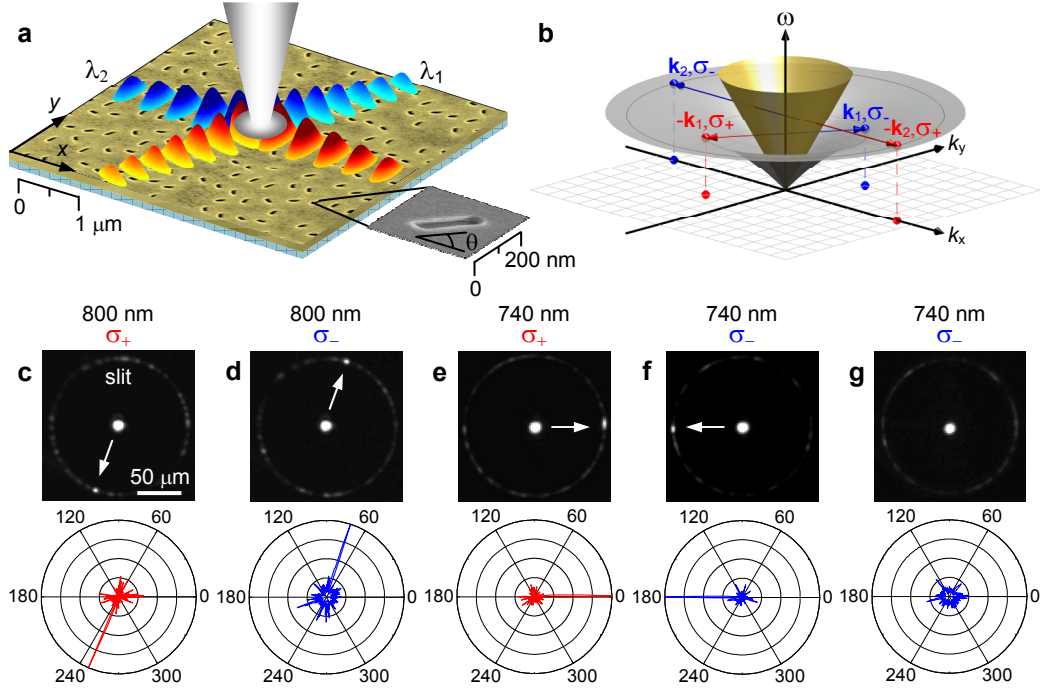

**Figure 1 | Near-field open channels via disordered metasurfaces. a**, Schematic of directional SPP open channels by a disordered metasurface. The scanning electron microscope image shows the $10 \times 10\,\mu m^2$ metasurface, wherein $r_{min} = 300\,nm$ and $d \approx 520\,nm$, fabricated using a focused ion beam. The array consists of 80-by-220-$nm^2$ nanoantennas, etched to a depth of 100 nm into a 200 nm thick gold film, evaporated onto a glass substrate. The diameter and width of the surrounding annular slit is 150 μm and 150 nm, respectively. **b**, Dispersion relation of free-space light (conic manifold) and SPPs (curved manifold). Red and blue arrows correspond to directional SPP coupling by normally incident light for $\sigma_\pm$ incident spin states, respectively. **c**,**d**, Measured intensities of open channels and corresponding azimuthal cross sections along the slit for $\sigma_\pm$ illuminations, respectively, at a wavelength of 800 nm. In the polar representation, the azimuthal angle is given in degrees and the intensity is on a linear scale. **e**,**f**, Measured intensities and azimuthal cross sections for $\sigma_\pm$, respectively, at a wavelength of 740 nm. **g**, Measured behavior of a metasurface with randomly oriented nanoantennas.

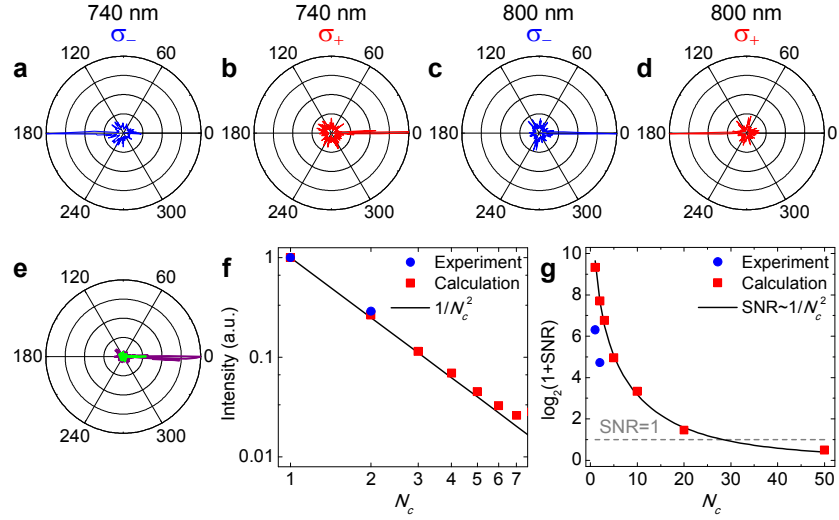

**Figure 2 | Information capacity analysis of multichannel disordered metasurface. a-d**, Spin-controlled open channels with similar directionality and different operating wavelengths. The measured intensities along the slit show the open channels for $\sigma_-$ and $\sigma_+$ excitations, at wavelengths of 740 (**a**,**b**) and 800 nm (**c**,**d**), respectively. **e**, Measured azimuthal cross sections of single (purple) and double (green) open channels. **f**, Dependence of the signal intensity on number of open channels. **g**, Number of bits per channel for varying number of open channels. The calculation was performed with a constant total number of antennas. Note that the experimental points follow the calculated trend, where the difference between them lies in the additional experimental noise source of the coupling slit.

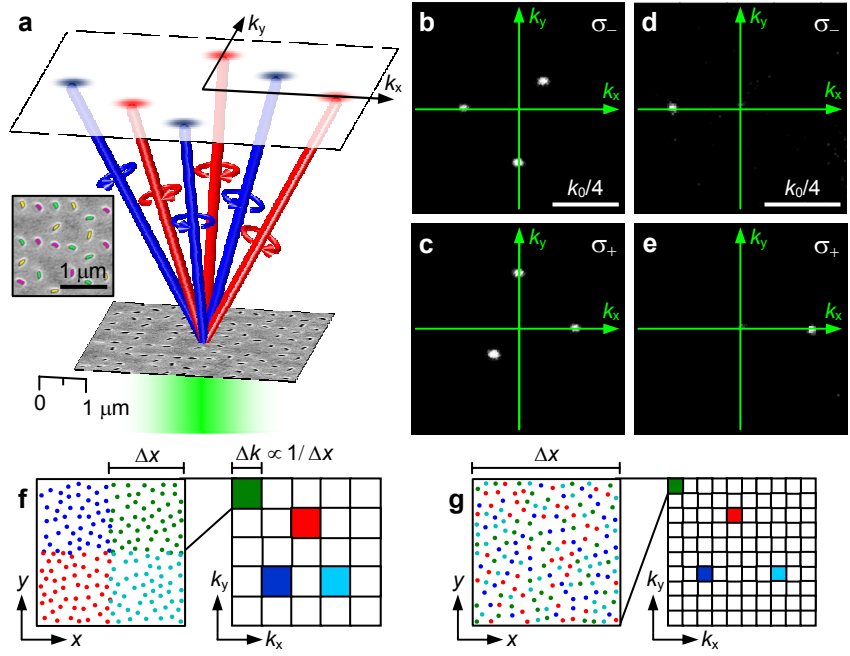

**Figure 3 | Free-space interconnects based on disordered metasurface. a**, Schematic of spin-controlled far-field open channels by a disordered metasurface. The inset shows the mixed antenna groups, where each color corresponds to a different channel. **b**,**c**, Spin-flip momentum deviations of 3 open channels for $\sigma_-$ and $\sigma_+$ spin states of the scattered light, respectively, at a wavelength of 740 nm. The polarization state is resolved with the use of a circular polarization analyzer (a quarter-wave plate followed by a linear polarizer). **d**,**e**, Spin-flip momentum deviations of a single open channel with $d \approx 2\lambda \approx 1.5\,\mu m$ for $\sigma_-$ and $\sigma_+$ scattered spin states, respectively. **f**,**g**, Metasurfaces divided into separated and mixed channel regions, respectively. The reciprocal spaces show that the channel capacity in the mixed channel type is significantly higher.